\documentstyle[prl,aps,preprint]{revtex}

\def\upleftarrow#1{\raise1.5ex\hbox{$\leftarrow$}\mkern-16.5mu #1}
\def\uprightarrow#1{\raise1.5ex\hbox{$\rightarrow$}\mkern-16.5mu #1}
\def\upleftrightarrow#1{\raise1.5ex\hbox{$\leftrightarrow$}\mkern-16.5mu #1}
\begin{document}
\tighten

\title{A QCD ANALYSIS OF\\
THE MASS STRUCTURE OF THE NUCLEON\thanks
{This work is supported in part by funds provided by the U.S.
Department of Energy (D.O.E.) under cooperative agreement
\#DF-FC02-94ER40818.}}

\author{Xiangdong Ji}

\address{Center for Theoretical Physics \\
Laboratory for Nuclear Science \\
and Department of Physics \\
Massachusetts Institute of Technology \\
Cambridge, Massachusetts 02139 \\
{~}   \\[3ex]
{\rm MIT-CTP: \#2368  \qquad\qquad\qquad\qquad   HEP-PH:  \#9410274}  \\
}

\date{Submitted to: {\it Physical Review Letters}
           \qquad\qquad   October 1994}

\maketitle

\begin{abstract}
{}From the deep-inelastic momentum sum rule
and the trace anomaly
of the energy-momentum tensor,
I derive a separation of the nucleon mass
into the contributions of the quark and gluon kinetic
and potential energies, the quark masses,
and the trace anomaly.
\end{abstract}

\narrowtext
The nucleon derives its mass (939 MeV) from
the quark-gluon dynamics of its underlying structure.
However, due to the complexity of the low-energy
Quantum Chromodynamics (QCD), a more detailed understanding
of the nucleon mass seems difficult.
The lattice QCD is successful in reproducing
the measured mass from the fundamental
lagrangian \cite{latt}, but the approach provides little insight
on how the number is partitioned between the nucleon's
quark and gluon content.
Years after the advent of QCD, our knowledge
of the nucleon's mass structure mostly comes from
models: non-relativistic quark models, Bag models,
the Skyrme model, string models, the Nambu-Jona-Lasinio model,
to just name a few. Though all the models are made to fit
the mass of the nucleon, they differ considerably
on the account of its origin. Depending on
different facets of QCD the models are created to
emphasize, the interpretations of the nucleon mass often
go opposite extremes.

In this Letter I show that an insight on the
mass structure of the nucleon can be produced within
QCD with the help of the deep-inelastic momentum
sum rule and the trace anomaly. The result is a separation
of the nucleon mass into the contributions from the quark, antiquark,
gluon kinetic and potential energies, the quark masses,
the gluon trace anomaly. Numerically, the only large uncertainty
is the size of $\langle P|m_s\bar ss|P \rangle$,
the strange scalar charge of the nucleon. Some implications of
this break-up of the masses are discussed following the result.

Let me begin with the energy-momentum tensor of QCD,
\begin{equation}
        T^{\mu\nu} = {1\over 2}\bar \psi i{\upleftrightarrow D}^{(\mu }
                  \gamma^{\nu)} \psi
               + {1\over 4}g^{\mu\nu}F^2
                - F^{\mu\alpha}F^\nu_{\ \alpha},
\label{tmn}
\end{equation}
where $\psi$ is the quark field with color, flavor, and Dirac indices;
$F^{\mu\nu}$ is the gluon field strength with color indices and
$F^2=F^{\alpha\beta}F_{\alpha\beta}$; and
all implicit indices are summed over. The covariant derivative
${\upleftrightarrow D}^{\mu} = {\uprightarrow D}^\mu- {\upleftarrow D}^\mu$,
with ${\uprightarrow D}^{\mu}={\uprightarrow \partial}^{\mu}
+igA^{\mu}$ and ${\upleftarrow D}^{\mu}={\upleftarrow \partial}^{\mu}
-igA^{\mu}$, where $A^{\mu} = A^{\mu}_at^a$ is the gluon potential.
The symmetrization of the indices $\mu$ and $\nu$ in the first
term is indicated by $(\mu\nu)$. Eq.~(\ref{tmn}) is
quite formal, for it contains neither
the gauge fixing and ghost terms, nor the
trace anomaly. The first type of terms are BRST-exact \cite{collins}
and have vanishing physical matrix elements according to
the Joglekar-Lee theorems \cite{jl}.
I will add the trace anomaly explicitly when the renormalization
issue is dealt with.

A few results about the energy-momentum tensor are well-known.
First of all, it is a symmetric and conserved tensor,
\begin{equation}
      T^{\mu\nu} = T^{\nu\mu}, \ \ \ \partial_\mu T^{\mu\nu} = 0.
\end{equation}
Because of the second property, the tensor is a finite operator
and does not need an overall renormalization.
All the fields and couplings in Eq.~(\ref{tmn}) are bare and
their divergence
are cancelled by the standard set of renormalization constants.
The only complication
is the gluon part of the tensor cannot be renormalized with a vanishing
trace (see below). Second, the tensor
defines the hamiltonian operator of QCD,
\begin{equation}
           H_{\rm QCD} = \int d^3\vec{x}~ T^{00}(0,\vec{x}),
\end{equation}
which is also finite and scale-independent.
Third, the matrix element of the tensor operator in
the nucleon state is \cite{jm},
\begin{equation}
          \langle P|T^{\mu\nu}|P \rangle = P^{\mu} P^{\nu}/M ,
\label{tmatrix}
\end{equation}
where $|P\rangle $ is the nucleon state with momentum
$P^{\mu}$ and is normalized according to $\langle P|P\rangle =
(E/M)(2\pi)^3\delta^3({\bf 0})$, and $E$ and $M$ is the energy and mass of
the nucleon, respectively. Lastly, the
trace of the tensor is \cite{anomaly}
\begin{equation}
       \hat  T^{\mu\nu} = {1\over  4}g^{\mu\nu}\Big[
         (1+\gamma_m)\bar \psi m\psi
         + {\beta(g) \over 2g}F^2\Big],
\label{trace}
\end{equation}
where $m$ is a quark mass matrix, $\gamma_m$ is the anomalous
dimension of the mass operator, and $\beta(g)$ is the $\beta$-function
of QCD. At the leading order $\beta(g) = -\beta_0 g^3/(4\pi)^2$ and
$\beta_0 = (11-2n_f/3)$, where $n_f$ is the number of
flavors. The second term is called the trace anomaly and is generated
in the process of renormalization.

According to the above, the mass of the nucleon is
\begin{equation}
      M = {\langle P |\int d^3\vec{x}~ T^{00}(0,\vec{x}) |P \rangle
             \over \langle P| P\rangle}
        \equiv  \langle T^{00} \rangle,
\label{mass}
\end{equation}
in the nucleon's rest frame.
Although I formally work with the matrix
elements of the nucleon, it actually is the difference
of the nucleon matrix elements and the vacuum matrix
elements that enters all the formula
(the vacuum has zero measurable energy density).
According to (\ref{mass}), a mass separation
can be found through
a decomposition of $T^{\mu\nu}$ into various
parts, which are then evaluated with the deep-inelastic
momentum sum rule and the scalar charge of the nucleon.
[Note that the parts of the energy-momentum tensor
are not separately conserved, so the breaking of
the nucleon energy cannot be Lorentz covariant.]

First of all, let me decompose the $T^{\mu\nu}$ into
traceless and trace parts,
\begin{equation}
         T^{\mu\nu} = \bar T^{\mu\nu} +  \hat T^{\mu\nu},
\label{dec11}
\end{equation}
where $\bar T^{\mu\nu}$ is traceless.
According to Eq.~(\ref{tmatrix}), I have,
\begin{eqnarray}
          \langle P|\bar T^{\mu\nu} |P \rangle&
       = &(P^\mu P^\mu - {1\over 4}M^2g^{\mu\nu})/M \label{less}, \\
          \langle P|\hat T^{\mu\nu} |P\rangle
            & = &{1\over 4}g^{\mu\nu}M.
\end{eqnarray}
Combining Eq.~(\ref{mass}) with the above three equations, I get,
\begin{eqnarray}
         \langle \bar T^{00} \rangle & = & {3\over 4} M,  \\
         \langle \hat T^{00} \rangle & = & {1\over 4} M. \label{dec12}
\end{eqnarray}
Thus 3/4 of the nucleon mass comes from the traceless part
of the energy-momentum
tensor and 1/4 from the trace part. The magic number 4 is just
the space-time dimension. This decomposition, a bit like the
virial theorem, is valid for any bound states in field theory!

The traceless part of the energy-momentum tensor can be decomposed into
the contribution from the quark and gluon parts,
\begin{equation}
          \bar T^{\mu\nu} = \bar T^{\mu\nu}_q + \bar T^{\mu\nu}_g,
\end{equation}
where
\begin{eqnarray}
          \bar T^{\mu\nu}_q & =& {1\over 2}\bar
         \psi i{\upleftrightarrow D}^{(\mu }\gamma^{\nu)} \psi
              - {1\over 4}g^{\mu\nu}\bar \psi m\psi,  \\
          \bar T^{\mu\nu}_g &=& {1\over 4}g^{\mu\nu}F^2
                - F^{\mu\alpha}F^\nu_{\ \alpha}.
\end{eqnarray}
Although the sum of $\bar T^{\mu\nu}_q$ and $\bar T^{\mu\nu}_g$
with bare fields and bare couplings
is finite
(now neglecting the trace anomaly),
 the individual operators are divergent and must be renormalized.
Under renormalization, they mix with each other and with other BRST-exact
and the equations of motion operators, which have vanishing physical matrix
elements \cite{collins,jl}.
For my purpose, I regard
both operators renormalized and dependent on a renormalization
scale $\mu^2$. Define their matrix elements in the nucleon state,
\begin{eqnarray}
          \langle P|\bar T_q^{\mu\nu}| P \rangle &=& a(\mu^2) (P^{\mu}P^{\nu}
               - {1\over 4}g^{\mu\nu}M^2)/M,     \label{defa}\\
         \langle P|\bar T_g^{\mu\nu}| P \rangle &=&
              (1-a(\mu^2)) (P^{\mu}P^{\nu}
               - {1\over 4}g^{\mu\nu}M^2)/M,
\end{eqnarray}
where I have used Eq.~(\ref{less}) to get the
second equation. The constant $a(\mu^2)$
is related to the deep inelastic sum rule \cite{yan},
\begin{equation}
        a(\mu^2) = \sum_f \int^1_0  x[q_f(x, \mu^2) + \bar q_f(x, \mu^2)]dx,
\end{equation}
where the sum is over all quark flavors and $q_f(x,\mu^2)$ and
$\bar q_f(x, \mu^2)$
are quark momentum distributions inside the nucleon in the infinite momentum
frame or light-front coordinate system. Again, according the
Eq.~(\ref{mass}),
I find the contribution to the nucleon mass,
\begin{eqnarray}
          \langle \bar T^{00}_q \rangle &=& {3\over 4} a(\mu^2) M,  \\
             \langle \bar T^{00}_g \rangle &= &{3\over 4} (1-a(\mu^2))M.
\end{eqnarray}

Finally, I turn to the trace part of the energy-momentum
tensor $\hat T^{\mu\nu}$. According to Eq.~(\ref{trace}),
I decompose it into $\hat T^{\mu\nu}_m$ and $\hat T^{\mu\nu}_a$,
the mass term and trace anomaly term,
respectively. Both operators are finite and scale independent.
If I define,
\begin{equation}
              b = 4 \langle \hat T^{00}_m \rangle/M, \label{defb}
\end{equation}
then according to Eq.~(\ref{dec12}), the anomaly part
contributes,
\begin{equation}
               \langle \hat T^{00}_a \rangle = {1\over 4} (1-b)M.
\end{equation}

Thus, the energy-momentum tensor $T^{\mu\nu}$ can be separated
into four gauge-invariant parts, $\bar T^{\mu\nu}_q$, $\bar T^{\mu\nu}_g$,
$\hat T^{\mu\nu}_m$, and $\hat T^{\mu\nu}_a$. They contribute,
respectively, $3a/4$, $3(1-a)/4$,
$b/4$, and $(1-b)/4$ fractions of the nucleon mass.
The corresponding breakdown for the hamiltonian is,
$H_{\rm QCD} = H_q' + H_g + H_m' + H_a,$ with
\begin{eqnarray}
           H_q' &=& \int d^3\vec{x}~\left[\bar
                                 \psi({-i{\bf D\cdot \alpha}})\psi
           + {3\over 4}\bar \psi m\psi\right],  \\
           H_g &=& \int d^3\vec{x}~ {1\over 2}
                                 ({\bf E}^2+{\bf B}^2),  \label{hg} \\
           H_m'&=& \int d^3\vec{x}~{1\over 4} \bar \psi m \psi,  \\
           H_a &=& \int d^3\vec{x}~{9\alpha_s\over 16\pi}
          ({\bf E}^2-{\bf B}^2). \label{ha}
\end{eqnarray}
where I have consistently neglected $\gamma_m$ and the higher-order terms in
$\beta(g)$. One can put them back if a higher precision analysis
becomes necessary. I also have taken $n_f=3$. [Note that the heavy
quarks do contribute to the mass term, the kinetic and potential energy
term, and the trace anomaly term. However, the contributions
cancel each other in the limit of $m_f\to \infty$, and thus for simplicity
I neglect them.]  If I rearrange the mass terms by defining,
\begin{eqnarray}
           H_q &= & \int d^3\vec{x}~\bar \psi({-i{\bf D\cdot \alpha}})\psi,
         \label{hq} \\
           H_m &=& \int d^3\vec{x}~\bar \psi m \psi, \label{hm}
\end{eqnarray}
then the QCD hamiltonian becomes,
\begin{equation}
    H_{\rm QCD}= H_q + H_m+ H_g +  H_a.
\end{equation}
Here $H_q$ (Eq.~(\ref{hq})) represents the quark and antiquark kinetic
and potential energies and contributes $3(a-b)/4$ fraction of the nucleon
mass. $H_m$ (Eq.~(\ref{hm})) is the quark mass term and contributes
$b$ fraction of the mass. $H_g$ (Eq.~(\ref{hg})) is
the normal part of the gluon energy and contributes $3(1-a)/4$ fraction
of the mass. Finally, $H_a$ (Eq.~(\ref{ha})) is the
gluon energy from the trace anomaly. It contributes $(1-b)/4$
fraction of the mass.

To determine the decomposition numerically, I
need the matrix elements $a$ and $b$.
The deep-inelastic scattering experiments
have determined $a(\mu^2)$ with an accuracy of
a few percent. Using a recent fit to the quark distributions \cite{cteq},
\begin{equation}
                 a_{{\overline{\rm MS}}}(1 {\rm GeV}^2) = 0.55
\end{equation}
where ${\overline {\rm MS}}$ refers to the modified minimal
subtraction scheme.

Without the heavy quarks, the matrix element $b$ is,
\begin{equation}
                  bM= \langle P|m_u\bar uu + m_d \bar dd |P \rangle
                        + \langle P |m_s \bar ss |P \rangle
\end{equation}
The first term is the $\pi N$ $\sigma$-term apart from
a small isospin-violating contribution of order 2 MeV\null.
A most recent analysis gave a magnitude of
45$\pm 5$ MeV for this term \cite{gl}. So the only unknown
in our analysis is the strange scalar charge $\langle P|m_s\bar ss|P \rangle$
in the nucleon. There are model calculations for this quantity in
the literature \cite{ha}. Here I choose to estimate it
using two standard approaches, though both of them are not
completely satisfactory.

In the first approach, the strange quark mass is considered
small in the QCD scale and so the chiral perturbation theory can be used to
calculate the SU(3) symmetry breaking effects. A
recent second-order analysis on the spectra of the baryon octet
combined with the measured $\sigma$-term yields \cite{gl},
\begin{eqnarray}
          \langle P|\bar ss|P\rangle & \simeq &
     0.11\times\langle P|\bar uu+\bar dd|P
           \rangle  \\
         & \simeq & 0.77,
\end{eqnarray}
where in the second line, I have used $(m_u+m_d)/2 \simeq 7$ MeV
at the scale of 1 GeV$^2$ \cite{na}. Taking the strange quark
mass to be 150 MeV at the same scale, I get,
\begin{equation}
            bM \simeq 160 \ {\rm MeV},
\end{equation}
Using Eq.~(\ref{dec12}), I have,
\begin{equation}
         \langle P|{\alpha_s\over \pi} F^2|P \rangle
            = - 693 \ {\rm MeV}.
\end{equation}

In the second approach, the strange quark is considered
heavy in the QCD scale. Using heavy-quark expansion,
it was found \cite{heavy1},
\begin{equation}
             \langle P|m_Q\bar QQ|P \rangle = - {1\over 12}\langle
P| {\alpha_s \over \pi}F^2|P\rangle.
\label{heavy}
\end{equation}
Thus the strange quark contribution in
\begin{equation}
       \langle P|\bar \psi m \psi + {\beta(g)\over 2g}F^2|P\rangle =  M,
\label{known}
\end{equation}
which is an explicit form of Eq.~(\ref{dec12}), cancels.
{}From the above equation and the $\sigma$-term,
I find,
\begin{equation}
         \langle P|{\alpha_s\over \pi} F^2|P \rangle
            = - 740 \ {\rm MeV}.
\end{equation}
This yields a strange matrix element
$\langle P|m_s\bar ss|P\rangle$ = 62 MeV\null.
Together with the $\sigma$-term, I determine,
\begin{equation}
            bM = 107 \ {\rm MeV}.
\end{equation}

The complete result of the mass decomposition at
the scale of $\mu^2=1 {\rm GeV}^2$,
together with the two numerical estimates,
is shown Table 1. I have not shown the errors
due to omission of higher-order perturbative effects and
errors on the $\sigma$-term and current quark masses.
The total effect on individual numbers is about 5 to 10
MeV\null. Thus I have rounded up the numbers to nearest 10 MeV\null.
The largest uncertainty is from the matrix element
$\langle P|m_s\bar ss|P\rangle$, which could be larger
than the difference of the two estimates shown. Nevertheless,
I will argue below that the total strange contribution
to the nucleon mass is quite small and with a smaller
uncertainty.

The following comments can be made with regard to
the numerical result.
\begin{itemize}
\item{The quark kinetic and potential energies
contribute about $1/3$ of the nucleon mass. Because the quark
kinetic energy must be very large when confined within
a radius of 1 fm, there must exists a large cancellation
between the kinetic and potential energies. This may not
be entirely surprising in the presence of strong interactions
between quarks and gluons. Such strong interactions are clearly
at the origin of the chiral symmetry breaking, modeled,
for instance, by the Nambu-Jona-Lasinio \cite{ha}.}
\item{The decomposition of the quark energy into different
flavors is possible. Taking the number 270 MeV (the $m_s\to 0$
limit)
as an example, I find the up-quark energy in the proton
is 250 MeV using the momentum fraction carried by up quark
0.375 \cite{cteq}, the down quark energy, 105 MeV, and strange quark
energy, $-85$ MeV\null. Further decomposition into valence and
sea contribution cannot be made without knowledge of
the separate valence and sea contributions to the scalar charge.}
\item{The quark mass term accounts for about 1/8 of the nucleon
mass. About half of which or more is carried by the strange
quark. The contributions from the up and down quarks
are well determined by the $\sigma$-term.}
\item{The normal gluon energy is about 1/3
of the nucleon mass and the trace anomaly part
contributes about 1/4. From these
two, I deduce the color-electric and color-magnetic
fields in the nucleon separately (take $\alpha_s$(1GeV)$\simeq 0.4$),
\begin{eqnarray}
      \langle P|{\bf E^2}|P \rangle & =& 1700 \ {\rm MeV}, \\
      \langle P|{\bf B^2}|P \rangle & =& -1050 \ {\rm MeV}.
\end{eqnarray}
So the magnetic-field energy is negative in the nucleon!
This of course is due to a cancellation between the quark's
magnetic field and  that of the vacuum. The electric field
in the vacuum is presumably small, however, it is large and positive
in the nucleon. This behavior of the color fields
in presence of quarks is very interesting,
it may help to unravel the structure of the QCD vacuum.}
\item{In the chiral limit, the gluon energy from the trace anomaly
($M/4$) corresponds exactly to the vacuum energy in the
MIT bag model \cite{bag}. The role of such energy in the model is to
confine quarks. Thus we see here a clear way through which
the scale symmetry breaking leads to quark confinement.
To keep the confinement mechanism, a model must
include $H_a$, the anomaly part of the hamiltonian.}
\item{The strange quark contributes about $-60$ MeV through
trace anomaly. When adding together with the kinetic and potential
energy contribution $-85$ MeV and the mass term 115 MeV (the
$m_s \to 0$ limit)
the total strange contribution to
the nucleon mass is a mere $-30$ MeV\null. (The other limit gives
a total of $-45$ MeV\null.) The smallness of the
contribution is, to a large extent, insensitive to the matrix element
$\langle P|m_s \bar ss|P\rangle$.}
\end{itemize}

To summarize, we have find a separation of the nucleon
mass into contributions from the quark kinetic
and potential energy, gluon energy, and the trace anomaly.
The largest uncertainty is from the strange matrix element
$\langle P|m_s \bar ss|P\rangle$. The result has interesting
implications on the quark-gluon structure of the nucleon
and on the response of the QCD vacuum to color charges.

I wish to thank D. Freedman, F. Low, and K. Johnson for useful discussions
and suggestions.

\begin{table}
\bigskip
\centering
\caption{A decomposition of the nucleon mass into different contributions.
The matrix elements $a$ and $b$ are defined in Eqs.~(\protect\ref{defa})
and (\protect\ref{defb}).}
\medskip
\label{tab1}
\mediumtext
\begin{tabular}{|@{\hspace{1.24em}}c@{\hspace{1.24em}}|
           @{\hspace{1.24em}}c@{\hspace{1.24em}}|
           @{\hspace{1.24em}}c@{\hspace{1.24em}}|
           @{\hspace{1.24em}}d@{\hspace{1.24em}}|
           @{\hspace{1.24em}}d@{\hspace{1.24em}}|}
{\rm mass type} & $H_i$ & $M_i$ & $m_s \to 0$(MeV) &
   $m_s \to \infty $(MeV)  \\ \hline
{\rm quark energy} & $\bar \psi({-i{\bf D\cdot \alpha}})\psi $ &
 $3(a-b)/4$ & 270  & 300  \\
{\rm quark mass} & $\bar \psi m\psi$ &
 $b$ & 160 &  110  \\
{\rm gluon energy} & ${1\over 2}({\bf E}^2+{\bf B}^2)$ &
 $3(1-a)/4$ & 320 &  320  \\
{\rm trace anomaly} & ${9\alpha_s\over 16\pi}({\bf E}^2-{\bf B}^2)$ &
 $(1-b)/4$ & 190 &  210 \\
\end{tabular}
\end{table}
\end{document}